\documentclass{article}

\usepackage{microtype}
\usepackage{graphicx}
\usepackage{subfigure}
\usepackage{booktabs} % for professional tables

\usepackage{hyperref}

\usepackage[accepted]{icml2024}

\usepackage{amsmath}
\usepackage{amssymb}
\usepackage{mathtools}
\usepackage{amsthm}
\usepackage{bm}

\usepackage[capitalize,noabbrev]{cleveref}

\theoremstyle{plain}

\theoremstyle{definition}

\theoremstyle{remark}

\usepackage[textsize=tiny]{todonotes}

\icmltitlerunning{ICML 2024 Machine Learning for Earth System Modeling workshop}

\begin{document}

\twocolumn[
% PH Version 1
\icmltitle{
A Generative Machine Learning Approach \\
 for Improving Precipitation from Earth System Models}

% It is OKAY to include author information, even for blind
% submissions: the style file will automatically remove it for you
% unless you've provided the [accepted] option to the icml2024
% package.

% List of affiliations: The first argument should be a (short)
% identifier you will use later to specify author affiliations
% Academic affiliations should list Department, University, City, Region, Country
% Industry affiliations should list Company, City, Region, Country

% You can specify symbols, otherwise they are numbered in order.
% Ideally, you should not use this facility. Affiliations will be numbered
% in order of appearance and this is the preferred way.
\icmlsetsymbol{equal}{*}

\begin{icmlauthorlist}
\icmlauthor{Philipp Hess}{tum,pik}
\icmlauthor{Niklas Boers}{tum,pik,exet}
\end{icmlauthorlist}

%\icmlaffiliation{tum}{Department of XXX, University of YYY, Location, Country}
\icmlaffiliation{tum}{School of Engineering \& Design, Earth System Modelling, Technical University Munich, Munich, Germany}
\icmlaffiliation{pik}{Potsdam Institute for Climate Impact Research, Potsdam, Germany}
\icmlaffiliation{exet}{Global Systems Institute and Department of Mathematics, University of Exeter, Exeter, UK}

\icmlcorrespondingauthor{Philipp Hess}{philipp.hess@tum.de}

% You may provide any keywords that you
% find helpful for describing your paper; these are used to populate
% the "keywords" metadata in the PDF but will not be shown in the document
\icmlkeywords{Machine Learning, ICML}

\vskip 0.3in
]

% this must go after the closing bracket ] following \twocolumn[ ...

% This command actually creates the footnote in the first column
% listing the affiliations and the copyright notice.
% The command takes one argument, which is text to display at the start of the footnote.
% The \icmlEqualContribution command is standard text for equal contribution.
% Remove it (just {}) if you do not need this facility.

%\printAffiliationsAndNotice{}  % leave blank if no need to mention equal contribution
\printAffiliationsAndNotice{\icmlEqualContribution} % otherwise use the standard text.

%PH:  only 4-6 sentences allowed
\begin{abstract}
Quantifying the impacts of anthropogenic global warming requires 
accurate Earth system model (ESM) simulations. 
Statistical bias correction and downscaling can be applied to reduce errors and increase the resolution of ESMs. 
However, existing methods, such as quantile mapping, cannot effectively improve spatial patterns or temporal dynamics.
We address this problem with a purely generative machine learning approach, combining unpaired domain translation with a super-resolution foundation model. 
Our results show realistic spatial patterns and temporal dynamics as well as reduced distributional biases in the processed ESM simulation.
\end{abstract}

\section{Introduction}

% impact modelling
Accurate projections of the impacts of anthropogenic climate change are of fundamental importance, e.g., in infrastructure planning, agriculture, or water resource management \cite{hatfield_climate_2011, wada_high-resolution_2016}. 
Extreme weather events, becoming more frequent and severe with increasing temperatures \cite{ipcc_climate_2021}, can have devastating environmental, economic, and societal effects \cite{mcmichael_insights_2012, kotz_effect_2022}.

Numerical Earth system models (ESMs) are our primary tool for projecting future climate scenarios. However, these models exhibit persistent biases, for which a main source are shortcomings in semi-empirical parameterizations of small-scale processes that cannot be resolved explicitly in ESMs \cite{schneider_climate_2017, balaji_are_2022}.
Such biases in ESMs are particularly problematic for climate impact models (CIMs), which translate large-scale atmospheric output from ESMs into local impacts, e.g., on urban flooding, landslides, or crop yield \cite{schreider_climate_2000, kang_climate_2009, gariano_landslides_2016}. 
Observational or reanalysis data is typically used to develop and calibrate CIMs. Hence, inconsistencies occur when ESM simulations -- e.g. caused by spatial biases, overly smooth spatial patterns, or unrealistic temporal correlations -- are provided as input instead \cite{wilby_comparison_1999}.

%Hence, accurate and high-resolution climate projections are indispensable to model impacts consistently and are crucial to informing policymakers and communities about the necessary mitigation and adaptation strategies.

Statistical downscaling and bias correction methods can efficiently correct biases in global ESM simulations. Quantile mapping (QM) \cite{panofsky_applications_1968, cannon_bias_2015, lange_trend-preserving_2019}, in particular, is often used to correct the entire distribution over all percentiles at the level of individual grid cells. However, QM can only be applied to univariate distributions at each location individually. Hence, it cannot effectively take spatial and inter-variable correlations into account.

Deep neural networks, on the other hand, can process high-dimensional data efficiently and have found many applications in weather prediction and climate modelling \cite{rasp_deep_2018, ravuri_skilful_2021, yuval_use_2021, hess_physically_2022, lam_learning_2023}. A unique challenge for bias correction and downscaling of ESMs with respect to observations is the unpaired training data \cite{hess_physically_2022, hess_deep_2023}. Due to the chaotic geophysical fluid dynamics, ESM fields will lose their pairing with the observational ground truth after a couple of simulated days, even if initiated on observation-based data \cite{lorenz_predictability_1995}.

%For downscaling, the A generative denoising diffusion model, trained on the unpaired ground truth dataset only that can later, at inference, be adapted to downscaling the ESM to higher resolution has been proposed (cite).

This study investigates the applicability of a purely generative machine learning approach that improves ESM simulations in a two-step manner: (i) bias correction at the native ESM resolution followed by (ii) downscaling to a higher resolution. \citealp{wan_debias_2023} have applied a similar strategy with different methods successfully to idealized fluid dynamical simulations.

For the bias correction (i), generative machine learning methods from unpaired image-to-image translation are a natural approach to correcting systematic errors in the ESM \cite{hess_physically_2022}.
While denoising diffusion models \cite{ho_denoising_2020, dhariwal_diffusion_2021, song_denoising_2022} are currently the state-of-the-art generative model for images, their Gaussian prior makes them challenging to apply for the unpaired domain translation task between two \emph{arbitrary} distributions \cite{torbunov_uvcgan_2023}. We therefore use cycle-consistent GANs \cite{zhu_unpaired_2017} for this task 
 \cite{pan_learning_2021, francois_adjusting_2021, hess_physically_2022, mcgibbon_global_2024}.
For downscaling (ii), we follow the SDEdit approach \cite{meng_sdedit_2022} using consistency models (CMs) \cite{song_consistency_2023} that can be trained on the observation-based target dataset only and adapted to the ESM later at inference, thereby bypassing the unpaired training data problem \cite{bischoff_unpaired_2023, hess_fast_2024}.
Combining CycleGAN and CM makes this novel two-step method highly efficient, requiring only a single network evaluation for each step. 

\section{Methods}
\subsection{Training Data}

 We use ESM simulations of daily precipitation from the fully coupled Potsdam Earth Model (POEM) \cite{druke_cm2mc-lpjml_2021}.
 As a ground truth, we use precipitation data from the ERA5 reanalysis \cite{hersbach_era5_2020} provided by the European Center for Medium-Range Weather Forecasting (ECMWF) with coverage from 1940 to present. 
 The preprocessing is done following \cite{hess_physically_2022, hess_fast_2024}, where the ERA5 data is bilinearly interpolated to the native POEM resolution of $3^\circ \times 3.75^\circ$, which corresponds to $60 \times 96$ pixel grey-scale images. 
 We use a 4 times higher resolution to train the downscaling model on a $0.75^\circ \times 0.9375^\circ$ grid, corresponding to $240 \times 384$ pixel images.
 
\subsection{Generative Adversarial Networks}

Generative adversarial networks \cite{goodfellow_generative_2014} learn a transformation $\mathbf{y} = g(\mathbf{x})$ of the sampling process $\mathbf{x} \sim p_x(\mathbf{x})$. Given two distributions of unpaired images $p_x(\mathbf{x})$ and $p_y(\mathbf{y})$, CycleGAN \cite{zhu_unpaired_2017} uses two \emph{generator} networks $g: X \mapsto Y$ and $f: Y \mapsto X$ to learn a translation between the two domains $X$ and $Y$ in an adversarial minimax game \cite{goodfellow_generative_2014}. For this purpose, two adversarial \emph{discriminators} networks $d_x: X \mapsto [0, 1]$ and $d_y: Y \mapsto [0,1]$, are used to classify whether a given sample is generated artificially or drawn from the target distribution. 
The generator $g(\cdot)$ then minimizes the loss 
\begin{align*}
    \mathcal{L}_g(\mathbf{x}, \mathbf{y})  = & \underbrace{\log [1- d_y(g(\mathbf{x}))]}_{\textrm{Adversarial loss}} \\
             &+ \lambda_{\mathrm{cyc}} \underbrace{|| g(f(\mathbf{y})) - \mathbf{y}||_1}_{\textrm{Cycle-consistency loss}} \\
             &+ \lambda_{\mathrm{idt}} \underbrace{|| g(\mathbf{y}) - \mathbf{y}||_1}_{\textrm{Identity loss} },
    \label{eq:cycle_gan}
\end{align*}
which includes an identity regularization to preserve content information in the source sample, and a cycle consistency loss. 
The latter computes the cycle translation error of a sample from the source to the target domain and back, enforcing a bijective mapping.

The discriminator $d_{y}(\cdot)$ maximizes the objective function
\begin{equation*}
\mathcal{L}_{d_y}(\mathbf{x},\mathbf{y}) = \log [d_y(\mathbf{y})]  + \log  [1- d_y(f(\mathbf{x}))].
\end{equation*}
The other generator $f(\cdot)$ and discriminator $d_x(\cdot)$ are trained analogously.
We follow the training strategy of \citealp{zhu_unpaired_2017} and use the same ResNet generator, PatchGAN discriminator, and hyperparameters as in \citealp{hess_physically_2022}.

\subsection{Consistency Models}

Consistency models (CMs) \cite{song_consistency_2023} are efficient emulators of denoising diffusion models. They learn a direct mapping from a Gaussian distribution at $t=t_{\mathrm{max}}$ to the target distribution $p_{\mathrm{data}}(\mathbf{x})$ at $t=t_{\mathrm{min}}$.
CMs only require a single step to generate a new sample but maintain a high degree of controllability at inference \cite{song_consistency_2023}. As the name suggests, CMs learn a \emph{self-consistent} function, parameterized by $\bm{\theta}$, 
\begin{equation}
c_{\bm{\theta}}(\mathbf{x}(t), t) = c_{\bm{\theta}}(\mathbf{x}(t'), t'), \; \forall \; t, t' \in [t_{\mathrm{min}}, t_{\mathrm{max}}].
\end{equation}
The training objective is given by 
\begin{equation*}
    \mathcal{L}_c \left( \mathbf{x} \right) = d \left(c_{\bm{\theta}}(\mathbf{\tilde{x}}_{n+1}, t_{n+1}),  c_{\bm{ \bar{\theta}} }\left(\mathbf{\tilde{x}}_{n}, t_{n} \right) \right), 
    \label{eq:loss}
\end{equation*}
where we denote $\mathbf{\tilde{x}}_{n+1} := \mathbf{x}+ t_{n+1}\mathbf{z}$, $\mathbf{\tilde{x}}_{n} := \mathbf{x}+ t_{n}\mathbf{z}$, 
 $\mathbf{x} \sim p_{\mathrm{data}}(\mathbf{x}), t_n \sim \mathcal{U}(1, N(k)-1)$, and $\mathbf{z} \sim \mathcal{N}(\mathbf{0}, \mathbf{1})$. The discretization $N(\cdot)$ increases over training time with a given schedule. We denote with $\bm{ \bar{\theta }}$ an exponential moving average (EMA) over the model parameters $\bm{\theta}$. 
The distance measure $d(\cdot, \cdot)$ combines the learned perceptual image patch similarity (LPIPS) \cite{zhang_unreasonable_2018} and the $l^1$ norm:
\begin{equation}
    d(\mathbf{x},\mathbf{y}) = \mathrm{LPIPS}(\mathbf{x},\mathbf{y}) + ||\mathbf{x}-\mathbf{y}||_1.
\end{equation}
We use the same UNet architecture \cite{ronneberger_u-net_2015} and training procedure as in \citealp{hess_fast_2024}. We choose a SDEdit noise time of $t^*=0.6$ from the intersection of the CycleGAN and POEM radially averaged power spectral densities, following \citealp{bischoff_unpaired_2023, hess_fast_2024}.

\subsection{Baseline}

Quantile mapping is used as a state-of-the-art statistical bias correction, which we combine with bilinear interpolation for downscaling \cite{lange_trend-preserving_2019}. QM estimates the cumulative distribution functions $\mathrm{CDF}_{\mathrm{obs}}$ and  $\mathrm{CDF}_{\mathrm{sim}}$ of the historical observations and ESM simulations, respectively. The bias correction is then performed with
\begin{equation}
    \hat{x}_{\mathrm{sim}}(t)= \mathrm{CDF}_{\mathrm{obs}}^{-1} (\mathrm{CDF}_{\mathrm{sim}}(x_{\mathrm{sim}}(t))),
\end{equation}
where $\hat{x}_{\mathrm{sim}}(t)$ is the correction at a given grid cell and time instance of the biased variable $x_{\mathrm{sim}}(t))$.

\begin{figure}[H]
\begin{center}
\centerline{\includegraphics[width=0.728\columnwidth]{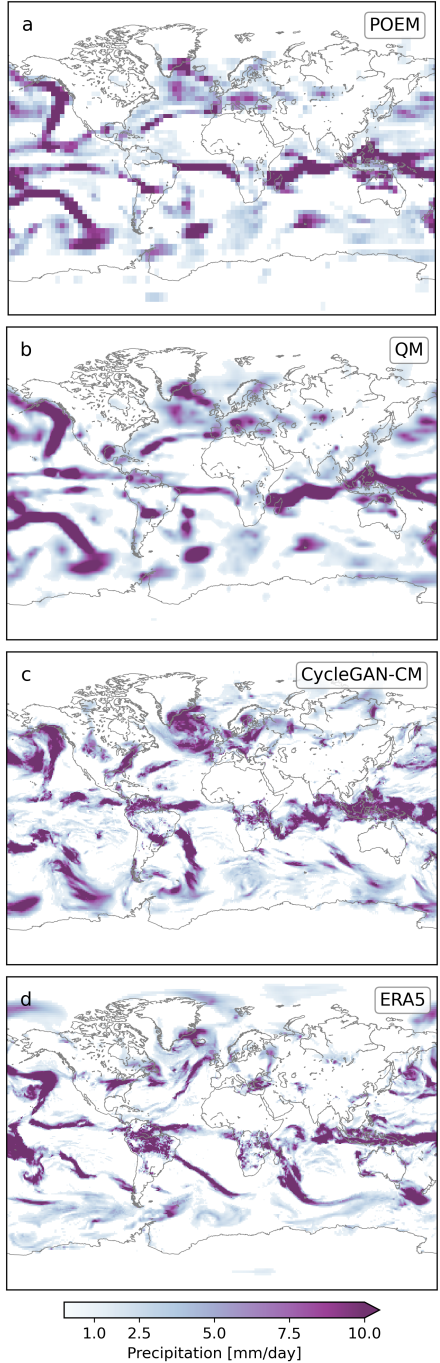}}
\caption{Qualitative comparison of single precipitation fields from the POEM ESM at its native resolution (a), the quantile mapping (QM) baseline (b), our CycleGAN-CM method (c), and a randomly selected, unpaired sample from the ERA5 ground truth (d). The CycleGAN-CM method shows realistic patterns across all spatial scales.}
\label{fig:single_fields}
\end{center}
\end{figure}
\section{Results}

% name test set.
\subsection{Spatial Fields}

% reslistic large and small scale
We qualitatively compare single spatial fields of daily precipitation from the POEM ESM, the two post-processing methods QM and CycleGAN-CM, and the ERA5 ground truth in Fig.~\ref{fig:single_fields}. 
Due to the low resolution, the POEM ESM is missing many fine-grained details that can be seen in the unpaired and four times higher resolution ERA5 target data. 
The QM method accurately preserves the large-scale patterns in the ESM, which are, however, too smooth compared to ERA5. Small-scale details are missing in the overly blurry fields. 
Our generative approach yields patterns that are qualitatively indistinguishable from the ERA5 ground truth across all spatial scales, while preserving the large-scale patterns from the ESM.

\begin{figure}[H]
\vskip 0.2in
\begin{center}
\centerline{\includegraphics[width=1.00\columnwidth]{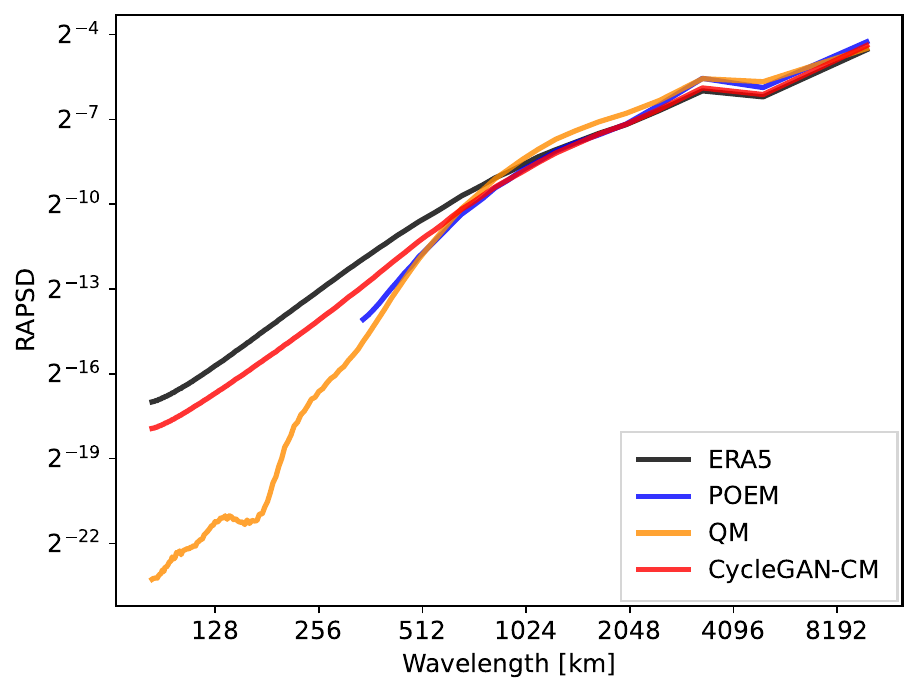}}
\caption{Radially averaged power spectral densities (RAPSDs) computed and averaged over single spatial fields for the ERA5 ground truth (black), the raw POEM ESM output (blue), the QM baseline (orange), and our CycleGAN-CM approach (red). Our method improves the RAPSD on all spatial scales.}
\label{fig:psd}
\end{center}
\vskip -0.2in
\end{figure}

Similar results are found in the radially averaged power spectral densities (RAPSDs) computed over the spatial fields in the ten-year test set from 2004-2014 (Fig.~\ref{fig:psd}). Towards smaller wavelengths, starting around 1000 km, the POEM ESM RAPSD underestimates the ERA5 target, while an overestimation at scales larger than 1000 km is visible. 
Applying QM together with bilinear interpolation to the ERA5 target resolution does not significantly improve the POEM RAPSD biases.
The generative approach combining GANs with CMs improves the RAPSD across all scales, and especially at the small wavelengths below 500 km. Interestingly, the RAPSD is also improved at large scales, which is not the case in \citealp{hess_fast_2024}, where QM instead of CycleGAN is used for bias correction.

\subsection{Temporal Correlations}

\begin{figure}[h]
\vskip 0.2in
\begin{center}
\centerline{\includegraphics[width=1.00\columnwidth]{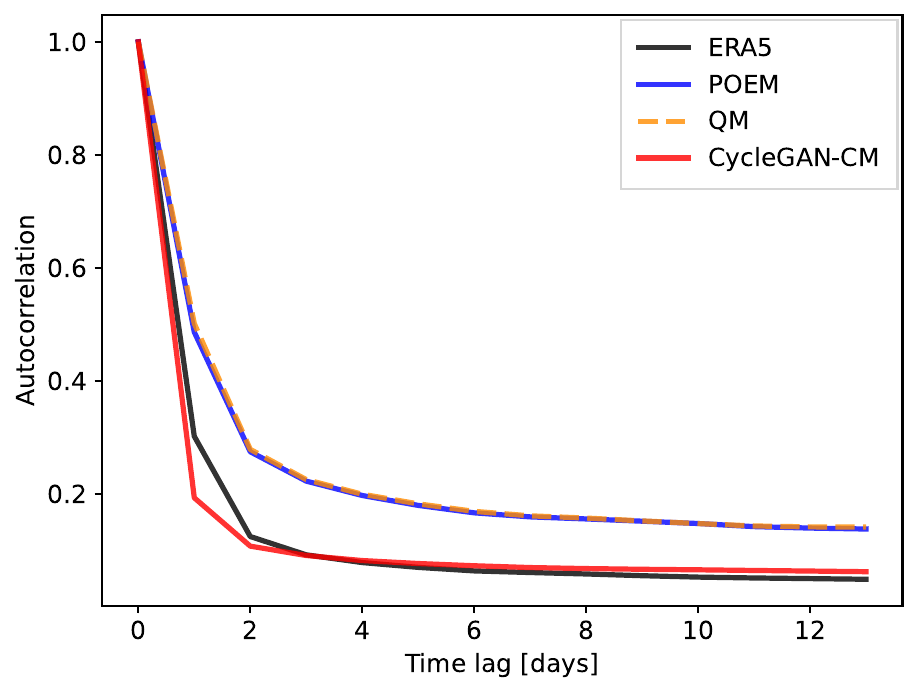}}
\caption{Autocorrelations computed at each grid cell in time and then averaged globally, for the ERA5 ground truth (black),  the POEM ESM (blue), the QM baseline (orange), and our CycleGAN-CM method (red). Our generative approach produces realistic temporal correlations.}
\label{fig:ac}
\end{center}
\vskip -0.2in
\end{figure}

Temporal correlations have often been overlooked in previous works on generative ML applications for bias correction and downscaling tasks. 
Yet temporal consistency is crucial for climate impact models that use the bias-corrected and downscaled ESM simulations as input.
Here, we report global means of the temporal autocorrelation (AC) with lags up to 14 days, computed for each grid cell in time (Fig.~\ref{fig:ac}).

The globally averaged AC for the POEM ESM simulations is too large for all time lags when compared to the ERA5 ground truth. This also agrees with the blurry spatial patterns in the ESM fields that lack the intermittency characteristic for spatial precipitation fields (Fig.~\ref{fig:single_fields}a).
A particularly strong overestimation of the AC can be seen in the tropics (Fig.~\ref{fig:ac_global}).
Applying QM results in temporal correlations nearly identical to the biased ESM, and hence no improvement.
Our stochastic generative approach shows a slight underestimation of the globally averaged AC for a one-day time lag (Fig.~\ref{fig:ac_global}), which can be primarily attributed to a pronounced underestimation over Antarctica (Fig.~\ref{fig:ac_global}).
Generally, our method shows an AC much closer to the ERA5 ground truth than the ESM and QM-processed simulation, which also agrees with the more intermittent and variable spatial fields in Fig.~\ref{fig:single_fields}c. 

\subsection{Bias Correction}
We evaluate the bias correction skill of our proposed generative approach using global histograms of the relative frequencies of events at a given magnitude, as well as latitude profiles of mean precipitation, following \citealp{hess_physically_2022}.

The POEM ESM strongly underestimates relative frequencies for extreme precipitation events in the right tail of the distribution (Fig.~\ref{fig:bias}a). 
QM and our CycleGAN-CM methods can improve the frequencies for extreme events with similar skills. We note that there is still room for improvement regarding the high-resolution ERA5 ground truth.

The double-peaked inter-tropical convergence zone (ITCZ) bias, which is common in many ESMs \cite{tian_double-itcz_2020}, is visible in the latitude-profile of mean precipitation from the POEM ESM (Fig.~\ref{fig:bias}b). QM removes the bias and closely agrees with the ERA5 ground truth, as expected.
The CycleGAN-CM method can also improve the long-term mean of precipitation while being trained on single daily fields only. The double-ITCZ is corrected with skill comparable to QM, with a slight underestimation of mean precipitation in the southern hemisphere.

\section{Discussion}

We propose a fully generative approach to bias-correct and downscale Earth system model simulations with respect to high-resolution reanalysis data. Unlike previous works \cite{fulton_bias_2023, hess_fast_2024}, our method does not include quantile mapping, which is unreliable in correcting spatial patterns and correlations.
Instead, the unpaired image-to-image translation is performed as a first step for bias correction at the coarse ESM resolution, using CycleGAN \cite{zhu_unpaired_2017}. A second network trained as a consistency model (CM) \cite{song_consistency_2023} on the target datasets is used to downscale the debiased fields to a high resolution. 

The here introduced CycleGAN-CM combination is more efficient than the similar approach in \citealp{wan_debias_2023}, requiring only a single network evaluation per task, which will be relevant when e.g. processing large ensemble simulations.

We evaluate temporal correlations, which are often overlooked in similar applications. Our method improves the autocorrelation to be more consistent with the ground truth. This positive results can be considered surprising, as we do not include temporal information in the network training. We believe this can be partially explained by the more intermittent spatial patterns generated with some stochasticity, which can be controlled with the noise variance of the downscaling \cite{hess_fast_2024}.

So far, we have only considered precipitation, one of the most challenging and important climate impact variables. However, our framework can be naturally extended to multiple variables by stacking fields as multiple color channels, which we leave for future work.

\section*{Software and Data}
The CycleGAN code is available on CodeOcean \url{https://doi.org/10.24433/CO.2750913.v1} and the consistency model on Github \url{https://github.com/p-hss/consistency-climate-downscaling.git}.
The reanalysis data can be obtained at the Copernicus Climate Change Service (C3S) 
(\url{https://cds.climate.copernicus.eu/cdsapp#!/dataset/reanalysis-era5-single-levels?tab=overview}).
The POEM ESM data is available for download at \url{https://doi.org/10.5281/zenodo.4683086}.

% Acknowledgements should only appear in the accepted version.
%\section*{Acknowledgements}

\section*{Impact Statement}

This work aims to improve Earth system model simulations of precipitation including extreme events. Extreme events can have a large impact on society causing, for example, floods and landslides, and are expected to become stronger and more frequent under ongoing climate change. 

\bibliography{references}

\begin{thebibliography}{41}
\providecommand{\natexlab}[1]{#1}
\providecommand{\url}[1]{\texttt{#1}}
\expandafter\ifx\csname urlstyle\endcsname\relax
  \providecommand{\doi}[1]{doi: #1}\else
  \providecommand{\doi}{doi: \begingroup \urlstyle{rm}\Url}\fi

\bibitem[Balaji et~al.(2022)Balaji, Couvreux, Deshayes, Gautrais, Hourdin, and Rio]{balaji_are_2022}
Balaji, V., Couvreux, F., Deshayes, J., Gautrais, J., Hourdin, F., and Rio, C.
\newblock Are general circulation models obsolete?
\newblock \emph{Proceedings of the National Academy of Sciences}, 119\penalty0 (47):\penalty0 e2202075119, November 2022.
\newblock \doi{10.1073/pnas.2202075119}.
\newblock URL \url{https://www.pnas.org/doi/10.1073/pnas.2202075119}.
\newblock Publisher: Proceedings of the National Academy of Sciences.

\bibitem[Bischoff \& Deck(2023)Bischoff and Deck]{bischoff_unpaired_2023}
Bischoff, T. and Deck, K.
\newblock Unpaired {Downscaling} of {Fluid} {Flows} with {Diffusion} {Bridges}, May 2023.
\newblock URL \url{http://arxiv.org/abs/2305.01822}.

\bibitem[Cannon et~al.(2015)Cannon, Sobie, and Murdock]{cannon_bias_2015}
Cannon, A.~J., Sobie, S.~R., and Murdock, T.~Q.
\newblock Bias {Correction} of {GCM} {Precipitation} by {Quantile} {Mapping}: {How} {Well} {Do} {Methods} {Preserve} {Changes} in {Quantiles} and {Extremes}?
\newblock \emph{Journal of Climate}, 28\penalty0 (17):\penalty0 6938--6959, September 2015.
\newblock ISSN 0894-8755, 1520-0442.
\newblock \doi{10.1175/JCLI-D-14-00754.1}.
\newblock URL \url{https://journals.ametsoc.org/view/journals/clim/28/17/jcli-d-14-00754.1.xml}.
\newblock Publisher: American Meteorological Society Section: Journal of Climate.

\bibitem[Dhariwal \& Nichol(2021)Dhariwal and Nichol]{dhariwal_diffusion_2021}
Dhariwal, P. and Nichol, A.
\newblock Diffusion {Models} {Beat} {GANs} on {Image} {Synthesis}, June 2021.
\newblock URL \url{http://arxiv.org/abs/2105.05233}.
\newblock arXiv:2105.05233 [cs, stat].

\bibitem[Drüke et~al.(2021)Drüke, von Bloh, Petri, Sakschewski, Schaphoff, Forkel, Huiskamp, Feulner, and Thonicke]{druke_cm2mc-lpjml_2021}
Drüke, M., von Bloh, W., Petri, S., Sakschewski, B., Schaphoff, S., Forkel, M., Huiskamp, W., Feulner, G., and Thonicke, K.
\newblock {CM2Mc}-{LPJmL} v1.0: biophysical coupling of a process-based dynamic vegetation model with managed land to a general circulation model.
\newblock \emph{Geoscientific Model Development}, 14\penalty0 (6):\penalty0 4117--4141, July 2021.
\newblock ISSN 1991-959X.
\newblock \doi{10.5194/gmd-14-4117-2021}.
\newblock URL \url{https://gmd.copernicus.org/articles/14/4117/2021/}.
\newblock Publisher: Copernicus GmbH.

\bibitem[François et~al.(2021)François, Thao, and Vrac]{francois_adjusting_2021}
François, B., Thao, S., and Vrac, M.
\newblock \emph{Adjusting spatial dependence of climate model outputs with cycle-consistent adversarial networks}.
\newblock Springer Berlin Heidelberg, 2021.
\newblock ISBN 0-12-345678-9.
\newblock \doi{10.1007/s00382-021-05869-8}.
\newblock URL \url{https://doi.org/10.1007/s00382-021-05869-8}.
\newblock Publication Title: Climate Dynamics Issue: 0123456789 ISSN: 14320894.

\bibitem[Fulton et~al.(2023)Fulton, Clarke, and Hegerl]{fulton_bias_2023}
Fulton, D.~J., Clarke, B.~J., and Hegerl, G.~C.
\newblock Bias {Correcting} {Climate} {Model} {Simulations} {Using} {Unpaired} {Image}-to-{Image} {Translation} {Networks}.
\newblock \emph{Artificial Intelligence for the Earth Systems}, 2\penalty0 (2), May 2023.
\newblock ISSN 2769-7525.
\newblock \doi{10.1175/AIES-D-22-0031.1}.
\newblock URL \url{https://journals.ametsoc.org/view/journals/aies/2/2/AIES-D-22-0031.1.xml}.
\newblock Publisher: American Meteorological Society Section: Artificial Intelligence for the Earth Systems.

\bibitem[Gariano \& Guzzetti(2016)Gariano and Guzzetti]{gariano_landslides_2016}
Gariano, S.~L. and Guzzetti, F.
\newblock Landslides in a changing climate.
\newblock \emph{Earth-Science Reviews}, 162:\penalty0 227--252, November 2016.
\newblock ISSN 0012-8252.
\newblock \doi{10.1016/j.earscirev.2016.08.011}.
\newblock URL \url{https://www.sciencedirect.com/science/article/pii/S0012825216302458}.

\bibitem[Goodfellow et~al.(2014)Goodfellow, Pouget-Abadie, Mirza, Xu, Warde-Farley, Ozair, Courville, and Bengio]{goodfellow_generative_2014}
Goodfellow, I., Pouget-Abadie, J., Mirza, M., Xu, B., Warde-Farley, D., Ozair, S., Courville, A., and Bengio, Y.
\newblock Generative {Adversarial} {Nets}.
\newblock In Ghahramani, Z., Welling, M., Cortes, C., Lawrence, N., and Weinberger, K.~Q. (eds.), \emph{Advances in {Neural} {Information} {Processing} {Systems}}, volume~27. Curran Associates, Inc., 2014.
\newblock URL \url{https://proceedings.neurips.cc/paper/2014/file/5ca3e9b122f61f8f06494c97b1afccf3-Paper.pdf}.

\bibitem[Hatfield et~al.(2011)Hatfield, Boote, Kimball, Ziska, Izaurralde, Ort, Thomson, and Wolfe]{hatfield_climate_2011}
Hatfield, J.~L., Boote, K.~J., Kimball, B.~A., Ziska, L.~H., Izaurralde, R.~C., Ort, D., Thomson, A.~M., and Wolfe, D.
\newblock Climate {Impacts} on {Agriculture}: {Implications} for {Crop} {Production}.
\newblock \emph{Agronomy Journal}, 103\penalty0 (2):\penalty0 351--370, 2011.
\newblock ISSN 1435-0645.
\newblock \doi{10.2134/agronj2010.0303}.
\newblock URL \url{https://onlinelibrary.wiley.com/doi/abs/10.2134/agronj2010.0303}.
\newblock \_eprint: https://onlinelibrary.wiley.com/doi/pdf/10.2134/agronj2010.0303.

\bibitem[Hersbach et~al.(2020)Hersbach, Bell, Berrisford, Hirahara, Horányi, Muñoz-Sabater, Nicolas, Peubey, Radu, Schepers, Simmons, Soci, Abdalla, Abellan, Balsamo, Bechtold, Biavati, Bidlot, Bonavita, De~Chiara, Dahlgren, Dee, Diamantakis, Dragani, Flemming, Forbes, Fuentes, Geer, Haimberger, Healy, Hogan, Hólm, Janisková, Keeley, Laloyaux, Lopez, Lupu, Radnoti, de~Rosnay, Rozum, Vamborg, Villaume, and Thépaut]{hersbach_era5_2020}
Hersbach, H., Bell, B., Berrisford, P., Hirahara, S., Horányi, A., Muñoz-Sabater, J., Nicolas, J., Peubey, C., Radu, R., Schepers, D., Simmons, A., Soci, C., Abdalla, S., Abellan, X., Balsamo, G., Bechtold, P., Biavati, G., Bidlot, J., Bonavita, M., De~Chiara, G., Dahlgren, P., Dee, D., Diamantakis, M., Dragani, R., Flemming, J., Forbes, R., Fuentes, M., Geer, A., Haimberger, L., Healy, S., Hogan, R.~J., Hólm, E., Janisková, M., Keeley, S., Laloyaux, P., Lopez, P., Lupu, C., Radnoti, G., de~Rosnay, P., Rozum, I., Vamborg, F., Villaume, S., and Thépaut, J.-N.
\newblock The {ERA5} global reanalysis.
\newblock \emph{Quarterly Journal of the Royal Meteorological Society}, 146\penalty0 (730):\penalty0 1999--2049, 2020.
\newblock ISSN 1477-870X.
\newblock \doi{10.1002/qj.3803}.
\newblock URL \url{https://onlinelibrary.wiley.com/doi/abs/10.1002/qj.3803}.
\newblock \_eprint: https://onlinelibrary.wiley.com/doi/pdf/10.1002/qj.3803.

\bibitem[Hess et~al.(2022)Hess, Drüke, Petri, Strnad, and Boers]{hess_physically_2022}
Hess, P., Drüke, M., Petri, S., Strnad, F.~M., and Boers, N.
\newblock Physically constrained generative adversarial networks for improving precipitation fields from {Earth} system models.
\newblock \emph{Nature Machine Intelligence}, 4\penalty0 (10):\penalty0 828--839, October 2022.
\newblock ISSN 2522-5839.
\newblock \doi{10.1038/s42256-022-00540-1}.
\newblock URL \url{https://www.nature.com/articles/s42256-022-00540-1}.
\newblock Number: 10 Publisher: Nature Publishing Group.

\bibitem[Hess et~al.(2023)Hess, Lange, Schötz, and Boers]{hess_deep_2023}
Hess, P., Lange, S., Schötz, C., and Boers, N.
\newblock Deep {Learning} for {Bias}-{Correcting} {CMIP6}-{Class} {Earth} {System} {Models}.
\newblock \emph{Earth's Future}, 11\penalty0 (10):\penalty0 e2023EF004002, 2023.
\newblock ISSN 2328-4277.
\newblock \doi{10.1029/2023EF004002}.
\newblock URL \url{https://onlinelibrary.wiley.com/doi/abs/10.1029/2023EF004002}.

\bibitem[Hess et~al.(2024)Hess, Aich, Pan, and Boers]{hess_fast_2024}
Hess, P., Aich, M., Pan, B., and Boers, N.
\newblock Fast, {Scale}-{Adaptive}, and {Uncertainty}-{Aware} {Downscaling} of {Earth} {System} {Model} {Fields} with {Generative} {Foundation} {Models}, March 2024.
\newblock URL \url{http://arxiv.org/abs/2403.02774}.
\newblock arXiv:2403.02774 [physics].

\bibitem[Ho et~al.(2020)Ho, Jain, and Abbeel]{ho_denoising_2020}
Ho, J., Jain, A., and Abbeel, P.
\newblock Denoising {Diffusion} {Probabilistic} {Models}, December 2020.
\newblock URL \url{http://arxiv.org/abs/2006.11239}.
\newblock arXiv:2006.11239 [cs, stat].

\bibitem[{IPCC}(2021)]{ipcc_climate_2021}
{IPCC}.
\newblock Climate {Change} 2021: {The} {Physical} {Science} {Basis}. {Contribution} of {Working} {Group} {I} to the {Sixth} {Assessment} {Report} of the {Intergovernmental} {Panel} on {Climate} {Change}.
\newblock Technical report, Cambridge University Press, 2021.
\newblock URL \url{https://www.ipcc.ch/report/sixth-assessment-report-working-group-i/}.

\bibitem[Kang et~al.(2009)Kang, Khan, and Ma]{kang_climate_2009}
Kang, Y., Khan, S., and Ma, X.
\newblock Climate change impacts on crop yield, crop water productivity and food security – {A} review.
\newblock \emph{Progress in Natural Science}, 19\penalty0 (12):\penalty0 1665--1674, December 2009.
\newblock ISSN 1002-0071.
\newblock \doi{10.1016/j.pnsc.2009.08.001}.
\newblock URL \url{https://www.sciencedirect.com/science/article/pii/S1002007109002810}.

\bibitem[Kotz et~al.(2022)Kotz, Levermann, and Wenz]{kotz_effect_2022}
Kotz, M., Levermann, A., and Wenz, L.
\newblock The effect of rainfall changes on economic production.
\newblock \emph{Nature}, 601\penalty0 (7892):\penalty0 223--227, January 2022.
\newblock ISSN 1476-4687.
\newblock \doi{10.1038/s41586-021-04283-8}.
\newblock URL \url{https://www.nature.com/articles/s41586-021-04283-8$}.
\newblock Number: 7892 Publisher: Nature Publishing Group.

\bibitem[Lam et~al.(2023)Lam, Sanchez-Gonzalez, Willson, Wirnsberger, Fortunato, Alet, Ravuri, Ewalds, Eaton-Rosen, Hu, Merose, Hoyer, Holland, Vinyals, Stott, Pritzel, Mohamed, and Battaglia]{lam_learning_2023}
Lam, R., Sanchez-Gonzalez, A., Willson, M., Wirnsberger, P., Fortunato, M., Alet, F., Ravuri, S., Ewalds, T., Eaton-Rosen, Z., Hu, W., Merose, A., Hoyer, S., Holland, G., Vinyals, O., Stott, J., Pritzel, A., Mohamed, S., and Battaglia, P.
\newblock Learning skillful medium-range global weather forecasting.
\newblock \emph{Science}, 382\penalty0 (6677):\penalty0 1416--1421, December 2023.
\newblock \doi{10.1126/science.adi2336}.
\newblock URL \url{https://www.science.org/doi/abs/10.1126/science.adi2336}.
\newblock Publisher: American Association for the Advancement of Science.

\bibitem[Lange(2019)]{lange_trend-preserving_2019}
Lange, S.
\newblock Trend-preserving bias adjustment and statistical downscaling with {ISIMIP3BASD} (v1.0).
\newblock \emph{Geoscientific Model Development}, 12\penalty0 (7):\penalty0 3055--3070, 2019.
\newblock ISSN 19919603.
\newblock \doi{10.5194/gmd-12-3055-2019}.

\bibitem[Lorenz(1995)]{lorenz_predictability_1995}
Lorenz, E.~N.
\newblock Predictability: a problem partly solved.
\newblock In \emph{Conference {Paper}}, Shinfield Park, Reading, 1995.
\newblock URL \url{https://www.ecmwf.int/node/10829}.

\bibitem[McGibbon et~al.(2024)McGibbon, Clark, Henn, Kwa, Watt-Meyer, Perkins, and Bretherton]{mcgibbon_global_2024}
McGibbon, J., Clark, S.~K., Henn, B., Kwa, A., Watt-Meyer, O., Perkins, W.~A., and Bretherton, C.~S.
\newblock Global {Precipitation} {Correction} {Across} a {Range} of {Climates} {Using} {CycleGAN}.
\newblock \emph{Geophysical Research Letters}, 51\penalty0 (4):\penalty0 e2023GL105131, 2024.
\newblock ISSN 1944-8007.
\newblock \doi{10.1029/2023GL105131}.
\newblock URL \url{https://onlinelibrary.wiley.com/doi/abs/10.1029/2023GL105131}.
\newblock \_eprint: https://onlinelibrary.wiley.com/doi/pdf/10.1029/2023GL105131.

\bibitem[McMichael(2012)]{mcmichael_insights_2012}
McMichael, A.~J.
\newblock Insights from past millennia into climatic impacts on human health and survival.
\newblock \emph{Proceedings of the National Academy of Sciences}, 109\penalty0 (13):\penalty0 4730--4737, March 2012.
\newblock \doi{10.1073/pnas.1120177109}.
\newblock URL \url{https://www.pnas.org/doi/abs/10.1073/pnas.1120177109}.
\newblock Publisher: Proceedings of the National Academy of Sciences.

\bibitem[Meng et~al.(2022)Meng, He, Song, Song, Wu, Zhu, and Ermon]{meng_sdedit_2022}
Meng, C., He, Y., Song, Y., Song, J., Wu, J., Zhu, J.-Y., and Ermon, S.
\newblock {SDEdit}: {Guided} {Image} {Synthesis} and {Editing} with {Stochastic} {Differential} {Equations}, January 2022.
\newblock URL \url{http://arxiv.org/abs/2108.01073}.
\newblock arXiv:2108.01073 [cs].

\bibitem[Pan et~al.(2021)Pan, Anderson, Goncalves, Lucas, Bonfils, Lee, Tian, and Ma]{pan_learning_2021}
Pan, B., Anderson, G.~J., Goncalves, A., Lucas, D.~D., Bonfils, C. J.~W., Lee, J., Tian, Y., and Ma, H.-Y.
\newblock Learning to {Correct} {Climate} {Projection} {Biases}.
\newblock \emph{Journal of Advances in Modeling Earth Systems}, 13\penalty0 (10):\penalty0 e2021MS002509, 2021.
\newblock ISSN 1942-2466.
\newblock \doi{10.1029/2021MS002509}.
\newblock URL \url{https://onlinelibrary.wiley.com/doi/abs/10.1029/2021MS002509}.
\newblock \_eprint: https://onlinelibrary.wiley.com/doi/pdf/10.1029/2021MS002509.

\bibitem[Panofsky(1968)]{panofsky_applications_1968}
Panofsky, H.~A.
\newblock Some applications of statistics to meteorology.
\newblock \emph{(No Title)}, 1968.
\newblock URL \url{https://cir.nii.ac.jp/crid/1130000797309178240}.

\bibitem[Rasp et~al.(2018)Rasp, Pritchard, and Gentine]{rasp_deep_2018}
Rasp, S., Pritchard, M.~S., and Gentine, P.
\newblock Deep learning to represent subgrid processes in climate models.
\newblock \emph{Proceedings of the National Academy of Sciences}, 115\penalty0 (39):\penalty0 9684--9689, September 2018.
\newblock ISSN 0027-8424, 1091-6490.
\newblock \doi{10.1073/pnas.1810286115}.
\newblock URL \url{http://www.pnas.org/lookup/doi/10.1073/pnas.1810286115}.

\bibitem[Ravuri et~al.(2021)Ravuri, Lenc, Willson, Kangin, Lam, Mirowski, Fitzsimons, Athanassiadou, Kashem, Madge, Prudden, Mandhane, Clark, Brock, Simonyan, Hadsell, Robinson, Clancy, Arribas, and Mohamed]{ravuri_skilful_2021}
Ravuri, S., Lenc, K., Willson, M., Kangin, D., Lam, R., Mirowski, P., Fitzsimons, M., Athanassiadou, M., Kashem, S., Madge, S., Prudden, R., Mandhane, A., Clark, A., Brock, A., Simonyan, K., Hadsell, R., Robinson, N., Clancy, E., Arribas, A., and Mohamed, S.
\newblock Skilful precipitation nowcasting using deep generative models of radar.
\newblock \emph{Nature}, 597\penalty0 (7878):\penalty0 672--677, 2021.
\newblock ISSN 14764687.
\newblock \doi{10.1038/s41586-021-03854-z}.
\newblock URL \url{http://arxiv.org/abs/2104.00954}.
\newblock arXiv: 2104.00954 Publisher: Springer US.

\bibitem[Ronneberger et~al.(2015)Ronneberger, Fischer, and Brox]{ronneberger_u-net_2015}
Ronneberger, O., Fischer, P., and Brox, T.
\newblock U-net: {Convolutional} networks for biomedical image segmentation.
\newblock In \emph{Lecture {Notes} in {Computer} {Science} (including subseries {Lecture} {Notes} in {Artificial} {Intelligence} and {Lecture} {Notes} in {Bioinformatics})}, volume 9351, pp.\  234--241. Springer Verlag, May 2015.
\newblock ISBN 978-3-319-24573-7.
\newblock \doi{10.1007/978-3-319-24574-4_28}.
\newblock URL \url{http://lmb.informatik.uni-freiburg.de/}.
\newblock arXiv: 1505.04597 ISSN: 16113349.

\bibitem[Schneider et~al.(2017)Schneider, Teixeira, Bretherton, Brient, Pressel, Schär, and Siebesma]{schneider_climate_2017}
Schneider, T., Teixeira, J., Bretherton, C.~S., Brient, F., Pressel, K.~G., Schär, C., and Siebesma, A.~P.
\newblock Climate goals and computing the future of clouds.
\newblock \emph{Nature Climate Change}, 7\penalty0 (1):\penalty0 3--5, January 2017.
\newblock ISSN 1758-6798.
\newblock \doi{10.1038/nclimate3190}.
\newblock URL \url{https://www.nature.com/articles/nclimate3190}.
\newblock Number: 1 Publisher: Nature Publishing Group.

\bibitem[Schreider et~al.(2000)Schreider, Smith, and Jakeman]{schreider_climate_2000}
Schreider, S.~Y., Smith, D.~I., and Jakeman, A.~J.
\newblock Climate {Change} {Impacts} on {Urban} {Flooding}.
\newblock \emph{Climatic Change}, 47\penalty0 (1):\penalty0 91--115, October 2000.
\newblock ISSN 1573-1480.
\newblock \doi{10.1023/A:1005621523177}.
\newblock URL \url{https://doi.org/10.1023/A:1005621523177}.

\bibitem[Song et~al.(2022)Song, Meng, and Ermon]{song_denoising_2022}
Song, J., Meng, C., and Ermon, S.
\newblock Denoising {Diffusion} {Implicit} {Models}, October 2022.
\newblock URL \url{http://arxiv.org/abs/2010.02502}.
\newblock arXiv:2010.02502 [cs].

\bibitem[Song et~al.(2023)Song, Dhariwal, Chen, and Sutskever]{song_consistency_2023}
Song, Y., Dhariwal, P., Chen, M., and Sutskever, I.
\newblock Consistency {Models}, May 2023.
\newblock URL \url{http://arxiv.org/abs/2303.01469}.
\newblock arXiv:2303.01469 [cs, stat].

\bibitem[Tian \& Dong(2020)Tian and Dong]{tian_double-itcz_2020}
Tian, B. and Dong, X.
\newblock The {Double}-{ITCZ} {Bias} in {CMIP3}, {CMIP5}, and {CMIP6} {Models} {Based} on {Annual} {Mean} {Precipitation}.
\newblock \emph{Geophysical Research Letters}, 47\penalty0 (8):\penalty0 1--11, 2020.
\newblock ISSN 19448007.
\newblock \doi{10.1029/2020GL087232}.

\bibitem[Torbunov et~al.(2023)Torbunov, Huang, Tseng, Yu, Huang, Yoo, Lin, Viren, and Ren]{torbunov_uvcgan_2023}
Torbunov, D., Huang, Y., Tseng, H.-H., Yu, H., Huang, J., Yoo, S., Lin, M., Viren, B., and Ren, Y.
\newblock {UVCGAN} v2: {An} {Improved} {Cycle}-{Consistent} {GAN} for {Unpaired} {Image}-to-{Image} {Translation}, September 2023.
\newblock URL \url{http://arxiv.org/abs/2303.16280}.
\newblock arXiv:2303.16280 [cs].

\bibitem[Wada et~al.(2016)Wada, de~Graaf, and van Beek]{wada_high-resolution_2016}
Wada, Y., de~Graaf, I. E.~M., and van Beek, L. P.~H.
\newblock High-resolution modeling of human and climate impacts on global water resources.
\newblock \emph{Journal of Advances in Modeling Earth Systems}, 8\penalty0 (2):\penalty0 735--763, 2016.
\newblock ISSN 1942-2466.
\newblock \doi{10.1002/2015MS000618}.
\newblock URL \url{https://onlinelibrary.wiley.com/doi/abs/10.1002/2015MS000618}.
\newblock \_eprint: https://onlinelibrary.wiley.com/doi/pdf/10.1002/2015MS000618.

\bibitem[Wan et~al.(2023)Wan, Baptista, Chen, Anderson, Boral, Sha, and Zepeda-Núñez]{wan_debias_2023}
Wan, Z.~Y., Baptista, R., Chen, Y.-f., Anderson, J., Boral, A., Sha, F., and Zepeda-Núñez, L.
\newblock Debias {Coarsely}, {Sample} {Conditionally}: {Statistical} {Downscaling} through {Optimal} {Transport} and {Probabilistic} {Diffusion} {Models}, May 2023.
\newblock URL \url{http://arxiv.org/abs/2305.15618}.
\newblock arXiv:2305.15618 [physics].

\bibitem[Wilby et~al.(1999)Wilby, Hay, and Leavesley]{wilby_comparison_1999}
Wilby, R.~L., Hay, L.~E., and Leavesley, G.~H.
\newblock A comparison of downscaled and raw {GCM} output: implications for climate change scenarios in the {San} {Juan} {River} basin, {Colorado}.
\newblock \emph{Journal of Hydrology}, 225\penalty0 (1):\penalty0 67--91, November 1999.
\newblock ISSN 0022-1694.
\newblock \doi{10.1016/S0022-1694(99)00136-5}.
\newblock URL \url{https://www.sciencedirect.com/science/article/pii/S0022169499001365}.

\bibitem[Yuval et~al.(2021)Yuval, O'Gorman, and Hill]{yuval_use_2021}
Yuval, J., O'Gorman, P.~A., and Hill, C.~N.
\newblock Use of {Neural} {Networks} for {Stable}, {Accurate} and {Physically} {Consistent} {Parameterization} of {Subgrid} {Atmospheric} {Processes} {With} {Good} {Performance} at {Reduced} {Precision}.
\newblock \emph{Geophysical Research Letters}, 48\penalty0 (6):\penalty0 1--11, 2021.
\newblock ISSN 19448007.
\newblock \doi{10.1029/2020GL091363}.
\newblock arXiv: 2010.09947.

\bibitem[Zhang et~al.(2018)Zhang, Isola, Efros, Shechtman, and Wang]{zhang_unreasonable_2018}
Zhang, R., Isola, P., Efros, A.~A., Shechtman, E., and Wang, O.
\newblock The {Unreasonable} {Effectiveness} of {Deep} {Features} as a {Perceptual} {Metric}, April 2018.
\newblock URL \url{http://arxiv.org/abs/1801.03924}.
\newblock arXiv:1801.03924 [cs].

\bibitem[Zhu et~al.(2017)Zhu, Park, Isola, and Efros]{zhu_unpaired_2017}
Zhu, J.-Y., Park, T., Isola, P., and Efros, A.~A.
\newblock Unpaired {Image}-{To}-{Image} {Translation} {Using} {Cycle}-{Consistent} {Adversarial} {Networks}.
\newblock In \emph{Proceedings of the {IEEE} {International} {Conference} on {Computer} {Vision}}, pp.\  2223--2232, 2017.
\newblock URL \url{https://openaccess.thecvf.com/content_iccv_2017/html/Zhu_Unpaired_Image-To-Image_Translation_ICCV_2017_paper.html}.

\end{thebibliography}
\bibliographystyle{icml2024}

%%%%%%%%%%%%%%%%%%%%%%%%%%%%%%%%%%%%%%%%%%%%%%%%%%%%%%%%%%%%%%%%%%%%%%%%%%%%%%%
%%%%%%%%%%%%%%%%%%%%%%%%%%%%%%%%%%%%%%%%%%%%%%%%%%%%%%%%%%%%%%%%%%%%%%%%%%%%%%%
% APPENDIX
%%%%%%%%%%%%%%%%%%%%%%%%%%%%%%%%%%%%%%%%%%%%%%%%%%%%%%%%%%%%%%%%%%%%%%%%%%%%%%%
%%%%%%%%%%%%%%%%%%%%%%%%%%%%%%%%%%%%%%%%%%%%%%%%%%%%%%%%%%%%%%%%%%%%%%%%%%%%%%%
\newpage
\appendix
\onecolumn
\section{Temporal correlations}

\begin{figure}[ht]
\vskip 0.2in
\begin{center}
\centerline{\includegraphics[width=0.7\columnwidth]{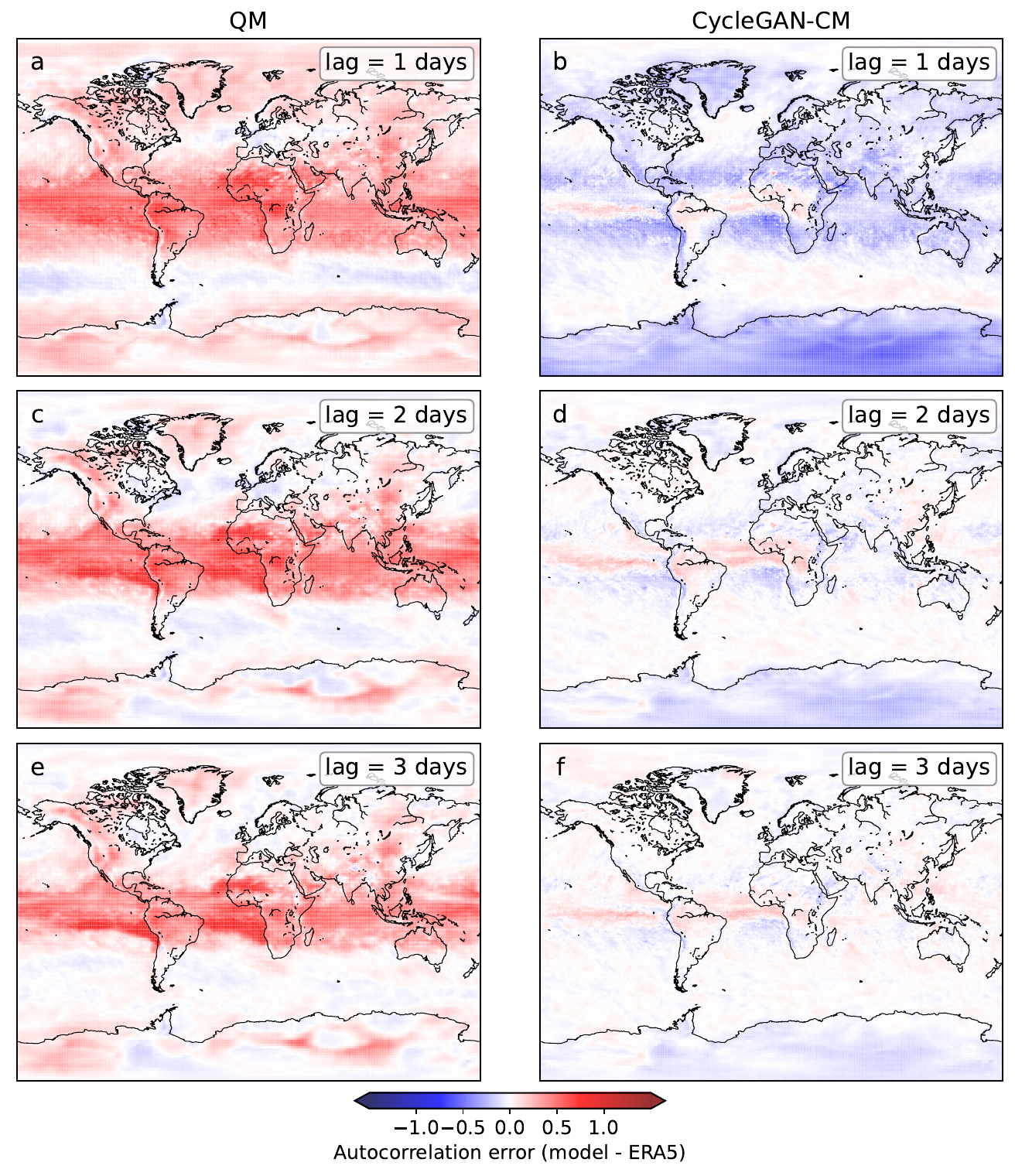}}
\caption{The difference error in autocorrelations for 1, 2, and 3-day lags are computed at each grid cell between ERA5 and the QM-processed (left column) or CycleGAN-CM-processed (right column) fields. A pronounced underestimation of autocorrelation is visible for 1-day time lags over the Antarctic continent of the CycleGAN-CM fields. The QM method shows a strong overestimation of autocorrelations in the tropics  for all time lags.} 
\label{fig:ac_global}
\end{center}
\vskip -0.2in
\end{figure}

%%%%%%%%%%%%%%%%%%%%%%%%%%%%%%%%%%%%%%%%%%%%%%%%%%%%%%%%%%%%%%%%%%%%%%%%%%%%%%%
%%%%%%%%%%%%%%%%%%%%%%%%%%%%%%%%%%%%%%%%%%%%%%%%%%%%%%%%%%%%%%%%%%%%%%%%%%%%%%%
\newpage
\section{Distributional biases}

\begin{figure}[ht]
\vskip 0.2in
\begin{center}
\centerline{\includegraphics[width=0.80\columnwidth]{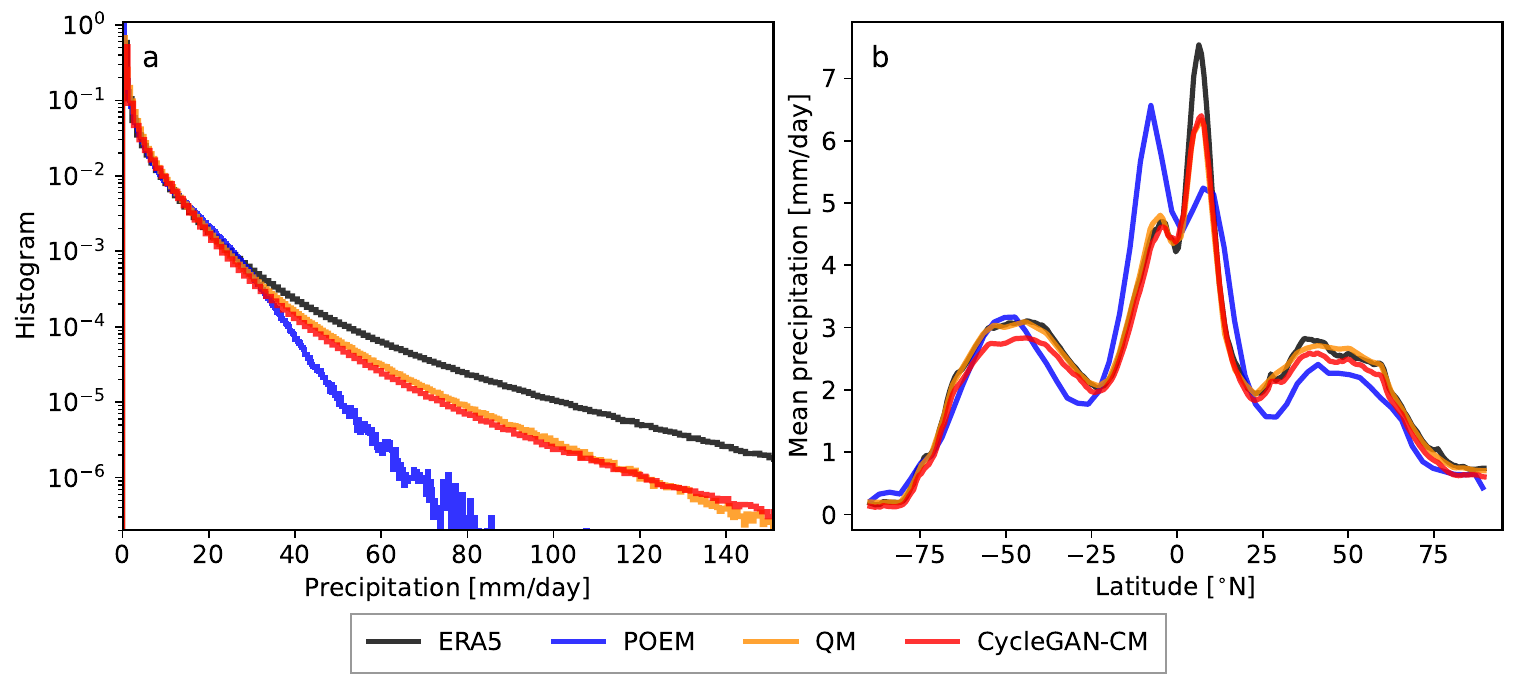}}
\caption{Comparison of relative frequency and longitude-mean biases. (a) Relative frequency histograms are shown for the ERA5 ground truth (black), the POEM ESM (blue), the QM-processed ESM (orange), and the CycleGAN-CM-processed ESM (red). (b) Precipitation averaged in time and longitude is shown for the same data as in (a).}
\label{fig:bias}
\end{center}
\vskip -0.2in
\end{figure}

\end{document}